\documentclass[letter,twocolumn]{jpsj3}
\usepackage{graphicx}
\usepackage{txfonts}
\usepackage{bm}
\usepackage{amsmath}
\usepackage{color}

\def\gsim{\mathop {\vtop {\ialign {##\crcr 
$\hfil \displaystyle {>}\hfil $\crcr \noalign {\kern1pt \nointerlineskip } 
$\,\sim$ \crcr \noalign {\kern1pt}}}}\limits}
\def\lsim{\mathop {\vtop {\ialign {##\crcr 
$\hfil \displaystyle {<}\hfil $\crcr \noalign {\kern1pt \nointerlineskip } 
$\,\,\sim$ \crcr \noalign {\kern1pt}}}}\limits}

\title{
Interplay of Valence and Semimetal-to-Insulator Transitions in SmS
}

\author{Shinji Watanabe}
\inst{ Department of Basic Sciences, Kyushu Institute of Technology, Kitakyushu, Fukuoka 804-8550, Japan
} 

\abst{
Recent discoveries of a new type of quantum criticality arising from Yb-valence fluctuations in Yb-based metal in periodic crystal and quasicrystal have opened a new class of quantum critical phenomena in correlated electron systems. To clarify whether this new concept can be generalized to other rare-earth-based semimetal and insulator, we study SmS which exhibits golden-black phase transition under pressure.
By constructing the model for SmS, we show that Coulomb repulsion between $4f$ and $5d$ orbitals at Sm drives first-order valence transition (FOVT) and semimetal-to-insulator transition (MIT) simultaneously, which explains the golden-black phase transition. We clarify the ground-state phase diagram for the FOVT and MIT by identifying the quantum critical point of the FOVT. We find that exciton condensates in both semimetal and insulator phases.
{Our result}
explains measured peak anomalies in the specific heat and compressibility in pressurized golden SmS and provides a cue to clarify recently-observed anomalies in black SmS.
}

\begin{document}
\maketitle


Quantum critical phenomena not following conventional magnetic criticality have attracted great interest in condensed matter physics. 
Recent discovery of the quantum critical point (QCP) of valence transition in the Yb-based periodic crystal~\cite{Kuga2018} and quasicrystal~\cite{Imura} 
opens a new class of quantum criticality originating from Yb valence fluctuations~\cite{WM2010}. 
So far, the valence QCP has been identified in the metallic Yb-based compounds. 
Then natural questions arise as: Are there any other systems for realizing the valence QCP? 
How does the valence transition occur in the semimetal and insulator?
To address these questions we focus on SmS as a typical system showing the valence transition and the semimetal-to-insulator transition (MIT). 

SmS has an NaCl type crystal structure of face-centered cubic. 
At ambient pressure, SmS is so-called Kondo insulator referred to as black phase ($b$-SmS). 
Under pressure above 6.5~kbar SmS undergoes isostructural phase transition from $b$-SmS to the golden phase ($g$-SmS) which is semimetal
accompanied by the valence transition as Sm$^{\nu+}$ from $\nu=2.0(2)$ to $2.67(4)$\cite{Coey1975,Deen,Imura2020}.
SmS has been studied for the last five decades~\cite{Jayaraman1970,RFK,Guseinov,Coey1975,Kaplan,Varma,Schweitzer,Egri,Iwamatsu,Kikoin,Wachter,Lehner,Antonov,Svane,Deen,Matsubayashi,Imura2009,Imura2011,Li,Kang2015,Imura2020}, but there remain several issues still unresolved. 
In $g$-SmS, a Schottkey-type anomaly was observed in the electric specific heat, thermal-expansion coefficient, and compressibility at low temperatures where 
the peak structure becomes prominent as pressure increases~\cite{Matsubayashi}. 
It is controversial whether $g$-SmS has a real gap or a pseudo gap~\cite{Wachter,Matsubayashi}. 
Furthermore, anomalies in thermal conductivity and Seebeck effect have recently been discovered in $b$-SmS at ambient pressure~\cite{Sato_PC}. 

To resolve these issues in $g$-SmS and also to get insight into $b$-SmS, we propose a simple model to describe the 
{golden-black}
phase transition in SmS. 
We clarify the ground state phase diagram for the valence transition and MIT by identifying the valence QCP. 
We show that exciton condensates in both semimetal and insulator phases  
{and our result}
explains measured anomalies in $g$-SmS under pressure and also gives a cue to clarify anomalies in $b$-SmS.

\begin{figure}[t]
\includegraphics[width=6.5cm]{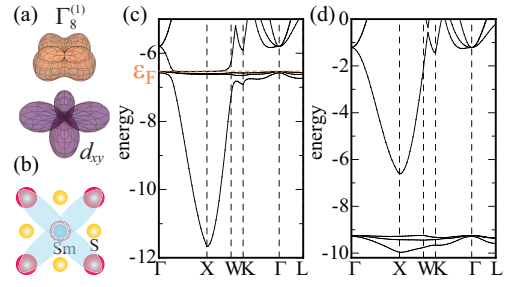}%
\caption{(color online)
(a) $\Gamma_{8}^{(1)}$ and $d_{xy}$ orbitals. 
(b) Schematic picture of exciton bound-state formation. 
Electric band structures $E_{e}$ for (c) $\varepsilon_f=-31.0$ 
and (d) $\varepsilon_f=-27.0$ at $U_{fd}=3.60$. 
Slater-Koster parameters are set as 
$(dd\sigma)=1$, $(dd\pi)=-0.3$, $(dd\delta)=-0.2$, $(ff\sigma)=-0.8$, $(ff\pi)=0.6$, $(ff\delta)=-0.4$, $(ffh)=0$, $(df\sigma)=0.9$, $(df\pi)=-0.4$, and $(df\delta)=0.3$. 
The ratio of 2nd (3rd) N.N. to N.N. transfers is set to be $-0.2$ (1.0) for $dd$ transfers, $0.8$ for $ff$ transfers and $0.125$ for $df$ hybridizations. 
In (c), $\varepsilon_{F}$ denotes the Fermi level. 
}
\label{fig:Ek}
\end{figure}

In the cubic system, the Sm $4f$ orbitals for the total angular momentum $J=5/2$ state are split into the $\Gamma_7$ doublet 
and $\Gamma_8$ quartet while the fivefold degenerate $5d$-orbitals at Sm are split into double degenerate $e_g$ and triply degenerate $t_{2g}$ orbitals. 
The band theory~\cite{Lehner,Antonov,Svane,Kang2015} and 
M\"{o}ssbauer spectroscopy studies~\cite{Coey1975} on SmS indicate that the physics is governed by 
valence fluctuations involving electrons of the $4f$ ($\Gamma_7$ and $\Gamma_8$) states and 
the conduction $t_{2g}$ states, 
$e^{-}+4f^5(\Gamma_7, \Gamma_8)\rightleftharpoons 4f^6$. 
The $\Gamma_7$ doublet and $\Gamma_8$ qualtet consist of the following orbitals: 
$|\Gamma_{7,\pm}\rangle=\sqrt{\frac{1}{6}}|\pm 5/2\rangle-\sqrt{\frac{5}{6}}|\mp 3/2\rangle$ and 
$|\Gamma^{(1)}_{8,\pm}\rangle=\sqrt{\frac{5}{6}}|\pm 5/2\rangle+\sqrt{\frac{1}{6}}|\mp 3/2\rangle$, 
$|\Gamma^{(2)}_{8,\pm}\rangle=|\pm 1/2\rangle$, respectively. 
This leads to the physical picture that the $f$ ($\Gamma_7$ and $\Gamma_8$) states hybridize with the $d$ $(t_{2g})$ states [see Fig.~\ref{fig:Ek}(a)], which 
describes the compensated metal with equal number of electrons and holes in $g$-SmS and the Kondo insulator in $b$-SmS. 


We now formulate our model for SmS partly following the model for SmB$_6$~\cite{Takimoto,ADC}. 
At each site, the $f$ and $d$ hole is described by an orbital and spin index, denoted by the combination $\zeta\equiv (a,\tilde{\sigma})$ 
{with 
$(a=\Gamma_8^{(1)}$, $\Gamma_8^{(2)}$, $\Gamma_7$, $\tilde{\sigma}=\eta=\pm)$ for $f$ states and $(a=xy, yz, zx, \tilde{\sigma}=\sigma=\uparrow,\downarrow)$ for $d$ states}.
The fields are given by the twelve component spinor
\[
\Psi_i=\left(
\begin{array}{c}
d_{\zeta}(i) \\
X_{0\zeta}(i)
\end{array}
\right)
\]
where $d_{\zeta}(i)$ destroys a $d$ hole at site $i$ and 
$X_{0\zeta}(i)=|4f^6\rangle\langle 4f^5,\zeta|$ is the Hubbard operator that destroys an $f$-hole at site $i$. 
The tight-binding Hamiltonian for the hybridized $d$-$f$ orbitals is 
\begin{eqnarray}
H_0=\sum_{\langle i,j\rangle}\Psi^{\dagger}_{i}h(i,j)\Psi_{j}
\end{eqnarray}
where the hopping matrix is given by
\[
h(i,j)= \left(
\begin{array}{cc}
h^{d}(i,j) & V(i,j) \\
V(i,j)^{\dagger} & h^{f}(i,j)
\end{array}
\right). 
\]
Here, the diagonal elements are transfers within $5d$ $(t_{2g})$ and $4f$ $(J=5/2)$ orbitals and the off diagonal elements are hybridization between them. 
To reproduce the local-density-approximation and dynamical mean-field theory (DMFT) band structures~\cite{Lehner,Kang2015}, we take transfers up to the third nearest neighbor (N.N.) Sm sites in $h^{d}(i,j)$ and
up to the second N.N. Sm sites in $h^{f}(i,j)$ and $V(i,j)$~\cite{Slater,Takegahara}. 

The Coulomb repulsion between $4f$ and $5d$ orbitals is 
\begin{eqnarray}
H_{U_{fd}}=U_{fd}\sum_{i\alpha\beta}X_{\alpha\alpha}(i)n^{d}_{i\beta}
+U'_{fd}\sum_{\langle i,j\rangle}\sum_{\alpha\beta}X_{\alpha\alpha}(i)n_{j\beta}^{d},
\label{eq:HUfd}
\end{eqnarray}
where $U_{fd} (U'_{fd})$ is onsite (inter site) Coulomb repulsion 
and $\langle i,j\rangle$ denotes the N.N. Sm sites. 
Here, $X_{\alpha\alpha}(i)=|4f^5,\alpha\rangle\langle 4f^5,\alpha|$ is the number operator of $f$ hole in the $\alpha=\Gamma_8^{(1)}, \Gamma_8^{(2)}$, and $\Gamma_7$ state and 
$n_{i\beta}^d=\sum_{\sigma}n_{i\beta\sigma}^d$ is the number operator of the $d$ hole defined by $n_{i\beta\sigma}^{d}\equiv d^{\dagger}_{i\beta\sigma}d_{i\beta\sigma}$ $(\beta=xy, yz, zx, \sigma=\uparrow, \downarrow)$. 
{Since other interactions such as the onsite and inter-site Coulomb repulsions between $5d$ orbitals are regarded to be included in the renormalized $t_{2g}$ bands, they are neglected in this study. }

Since the strong Coulomb repulsion affects each $f$ orbital we consider the following Hamiltonian 
{in the hole picture}
within the restriction of prohibiting the doubly-occupied $f$ orbitals: 
\begin{eqnarray}
H=H_0+H_{U_{fd}}.
\label{eq:H}
\end{eqnarray}
To analyze the ground-state and finite-temperature properties of this model, 
we employ a slave boson formulation of the Hubbard operators~\cite{Takimoto,ADC}, writing $X_{\zeta 0}(i)=f^{\dagger}_{i\zeta}b_i$, where $f_{\zeta}^{\dagger}|0\rangle\equiv |4f^5,\zeta\rangle$ creates an $f$ hole in the $\Gamma_8$ quartet and $\Gamma_7$ doublet while $b^{\dagger}|0\rangle\equiv|4f^6\rangle$ denotes the singlet filled $4f$ shell, subject to the constraint 
$Q_i=b_i^{\dagger}b_i+\sum_{\zeta}f^{\dagger}_{i\zeta}f_{i\zeta}=1$ at each site. 
We use the mean-field treatment by replacing the slave-boson operator $b_i$ by its expectation value $\bar{b}=\langle b_i\rangle$. 
The total $f(d)$-hole number is given by 
$\bar{n}_f=\sum_{\alpha\eta}\langle n^{f}_{i\alpha\eta}\rangle$ $(\bar{n}_d=\sum_{\beta}\langle n^d_{i\beta}\rangle)$ with $n_{i\alpha\eta}^{f}\equiv f^{\dagger}_{i\alpha\eta}f_{i\alpha\eta}$ where $\eta=\pm$ denote the Kramers states. 

For $H_{U_{fd}}$, we apply the mean-field approximation. 
The first term in Eq.~(\ref{eq:HUfd}) i.e. on site term is approximated as
$
U_{fd}\sum_{i\alpha\beta}
[
\langle X_{\alpha\alpha}(i)\rangle n^{d}_{i\beta}$$+X_{\alpha\alpha}(i)\langle n^{d}_{i\beta}\rangle
]
$. 
The second term in Eq.~(\ref{eq:HUfd}) is approximated as 
$
U'_{fd}\sum_{\langle i,j\rangle\alpha\beta}
[
\langle X_{\alpha\alpha}(i)\rangle n^{d}_{j\beta}$$+X_{\alpha\alpha}(i)\langle n^{d}_{j\beta}\rangle
]-U'_{fd}\sum_{\langle i,j\rangle\alpha\eta\beta\sigma}
[
\langle X_{\alpha\eta 0}(i)d_{j\beta\sigma}\rangle d^{\dagger}_{j\beta\sigma}X_{0\alpha\eta}(i)
+X_{\alpha\eta 0}(i)d_{j\beta\sigma}\langle d^{\dagger}_{j\beta\sigma}X_{0\alpha\eta}(i)\rangle
]
$. 

A possibility of the formation of the exciton, which is a bound state of an electron and a hole, in semimetal and semiconductor has been discussed theoretically~\cite{RFK,Guseinov,Schweitzer,Egri,Iwamatsu,Kikoin,Halperin,Kasuya,Otsuki,Kaneko,
Mazza}. The exciton formation by $H_{U_{fd}}$ in semimetal is intuitively understandable if we apply the hole-to-electron transformation to the $d$ orbital in Eq.~(\ref{eq:H}), which makes the sign change of $U_{fd}$ and $U'_{fd}$ in Eq.~(\ref{eq:HUfd}). Namely, the attractive force between an electron and a hole in semimetal gives rise to the exciton formation. This has indeed been confirmed by the microscopic theoretical calculation in the impurity model with the $U_{fd}$ term~\cite{Otsuki}. 
In SmS, 
the bound state of an electron and a hole is expected to be formed in the vicinity of Sm~\cite{Kikoin,Kasuya,Matsubayashi,Imura2011} [see Fig.~\ref{fig:Ek}(b)]. 
Hence we consider the exciton condensation characterized by the order parameter $\Delta_{\alpha\eta,\beta\sigma}(i,j)\equiv
{U'_{fd}}
\langle d^{\dagger}_{j\beta\sigma}X_{0\alpha\eta}(i)\rangle$. 
We consider the uniform condensation 
\begin{eqnarray}
\Delta_{\alpha\eta,\beta\sigma}(\xi)=\frac{U_{fd}'}{N}\sum_{\bm k}e^{i{\bm k}\cdot{\bm r}_{\xi}}\langle d^{\dagger}_{{\bm k}\beta\sigma}f_{{\bm k}\alpha\eta}\rangle\bar{b}
\label{eq:Dlt}
\end{eqnarray}
with ${\bm r}_{\xi}\equiv {\bm r}_{j}-{\bm r}_{i}$. 
The exciton formation by $\alpha\eta$$=\Gamma_{8+(-)}^{(1)}$ hole and $\beta\sigma$$=xy$$\downarrow$$(\uparrow)$ electron at the N.N. Sm sites is considered to be most probable 
since these orbitals give the largest hybridization between the N.N. Sm sites so that $U'_{fd}$ is most effective [see Fig.~\ref{fig:Ek}(a)]. 
{It is noted that some symmetry breaking can occur by exciton condensation~\cite{Mazza}. However, so far any change of the crystal symmetry from the space group $Fm\bar{3}m$ (No. 225) has not been reported on cooling in SmS. Hence, as a first step of analysis, w}e 
analyze the effect of the isotropic condensation as 
$\Delta\equiv\Delta_{\Gamma_{8}^{(1)}+,xy\downarrow}(\xi)=\Delta^{*}_{\Gamma_{8}^{(1)}-,xy\uparrow}(\xi)$
for the N.N. Sm sites ${\bm r}_{\xi=1\sim 12}\in (\pm a/2, \pm a/2,0)$, $(0, \pm a/2, \pm a/2)$, $(\pm a/2, 0, \pm a/2)$ [see Fig.~\ref{fig:Ek}(b)]. 
{Since this state does not break symmetry of the Hamiltonian, $\Delta$ is not the order parameter in strict sense. Hence, we refer to $\Delta$ as exciton condensation (EC) parameter hereafter. }


We solve the mean-field equations at half filling $\bar{n}= \bar{n}_f+\bar{n}_d=6$ 
for $\bar{b}$, Lagrange multiplayer $\lambda$ to require the constraint $\langle Q_i\rangle=1$, $\bar{n}_d$, Re$\Delta$, Im$\Delta$, and the chemical potential $\mu$ 
self-consistently. 
As typical values, we set $U'_{fd}=U_{fd}/10$ and the energy difference between $\Gamma_8$ qualtet and $\Gamma_7$ doublet $\Delta\varepsilon_f=0.04$ with 
$\varepsilon_{\Gamma_{7}}\equiv\varepsilon_f+\Delta\varepsilon_f$
where $f$ level is defined by the energy of the $\Gamma_8$ qualtet as $\varepsilon_f\equiv\varepsilon_{\Gamma^{(1)}_8}=\varepsilon_{\Gamma^{(2)}_8}$. 
{Here the energy unit is taken as the Slater-Koster parameter $(dd\sigma)$ and also hereafter.}
Numerical calculations are performed in the $N=32^3$ lattices for most cases. The band structures shown in Figs.~\ref{fig:Ek}(c) and \ref{fig:Ek}(d) are calculated in the $N=64^3$ lattice.

The electric band structure $E_{e}\equiv -E$ for $\varepsilon_f=-31.0$ and $U_{fd}=3.60$ at half filling $\bar{n}\equiv\bar{n}_f+\bar{n}_d=6$ is shown in Fig.~\ref{fig:Ek}(c). 
{Here, $E$ is the energy band in the hole picture.}
The hybridized flat bands with Sm $f$ character are formed around $E_{e}\sim -6.6$, which has a maximum at the $\Gamma$ point while the dispersive bands with Sm $d$ character have a minimum at the X point. 
The Fermi level $\varepsilon_{F}\equiv-\mu(T=0)$ is located at the hybridized bands, which indicates semimetal since the present half-filled state is the compensated metal with the equal number of electrons and holes. 
{Namely, the hole band and the electron band appear above and below $\varepsilon_F$ respectively.}
This overall feature is similar to the DMFT band structure of $g$-SmS~\cite{Kang2015}. 

On the other hand, for $\varepsilon_f=-27.0$ and $U_{fd}=3.60$ at half filling, the lower three bands are filled and upper three bands are empty as shown in Fig.~\ref{fig:Ek}(d), indicating the insulator with the energy gap, 
 which are also analogous to the DMFT band structure of $b$-SmS~\cite{Kang2015}. 

For deep $\varepsilon_f$, $\bar{n}_f=1$ is realized since the energy gain by occupied $f$ level overcomes the energy loss by $U_{fd}$. 
As $\varepsilon_f$ increases, the energy gain decreases and at a threshold $\varepsilon_f=\varepsilon^{v}_f$ the $f$ occupancy is expected to drop suddenly 
since large $U_{fd}$ forces a hole to pour into either the $f$ or $d$ orbital. 
On the contrary, for $U_{fd}=0$, the $f$ occupancy continuously changes as a function of $\varepsilon_f$. 
Hence, we expect that a critical value of $U_{fd}$ between the first-order transition and crossover exists in the ground state. 

\begin{figure}[t]
\includegraphics[width=6.5cm]{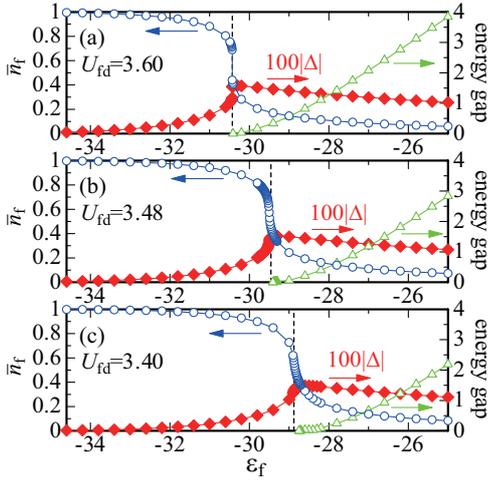}%
\caption{(color online) The $\varepsilon_f$ dependence of $\bar{n}_f$ (circle, left axis), energy gap (triangle, right axis), and $100|\Delta|$ (filled diamond, right axis)
for (a) $U_{fd}=3.60$, (b) 3.48, and (c) 3.40. Dashed lines are guides for the eyes at (a) $\varepsilon_f=\varepsilon_f^{v}$, (b) $\varepsilon_f^{QCP}$, and (c) $\varepsilon_f^{vc}$. 
}
\label{fig:nf_gap_Dlt}
\end{figure}

The effect of the $f$-level change is examined in Fig.\ref{fig:nf_gap_Dlt}(a) for $U_{fd}=3.60$.
As $\varepsilon_f$ increases, $\bar{n}_f$ decreases and shows a sudden drop at $\varepsilon_f=-30.43\equiv \varepsilon_f^{v}$. 
This indicates that the FOVT takes place at $\varepsilon_f=\varepsilon_f^{v}$. 
We find that the energy gap defined by the difference between the bottom of the conduction band and the top of the valence band starts to open at $\varepsilon_f=\varepsilon^{v}_f$. These results indicate that $U_{fd}$ drives the MIT 
{and}
the FOVT simultaneously.
At $\varepsilon_f=\varepsilon^{v}_f$, the energy balance $\tilde{\varepsilon}_f\approx\mu$ holds in the mean-field picture with $\tilde{\varepsilon}_f$ being the renormalized $f$ level $\tilde{\varepsilon}_f\equiv\varepsilon_f+\lambda+U_{fd}\bar{n}_d+6U'_{fd}\bar{n}_d$, 
where $\mu$ is located just at the point where the bottom of the conduction band touches the top of the valence band. 
As further $\varepsilon_f$ increases from $\varepsilon^{v}_f$, $\tilde{\varepsilon}_f$ rises since $\bar{n}_d$ increases as the result of charge transfer from the $f$ orbital caused by $U_{fd}$. This results in the shift down of the fully filled valence bands in the electron picture [e.g. three bands located around $E_{e}=-\tilde{\varepsilon}_f=-9.25$ in Fig~\ref{fig:Ek}.(d)]. Namely, to earn the energy gain of the charge-transfer-induced shift down of the filled valence bands, the energy gap opens for $\varepsilon_f>\varepsilon_f^{v}$.

For smaller $U_{fd}$, the magnitude of the $\bar{n}_f$ jump at the FOVT diminishes. 
At $U_{fd}=3.48$, the jump in $\bar{n}_f$ vanishes and the slope $-\partial\bar{n}_f/\partial \varepsilon_f\equiv\chi_v$ diverges at $\varepsilon_f=-29.46$, as shown 
in Fig.~\ref{fig:nf_gap_Dlt}(b). 
Here, $\chi_v$ is the valence susceptibility and the QCP of the FOVT is identified to be $(\varepsilon_f^{QCP}, U_{fd}^{QCP})=(-29.46, 3.48)$ where the critical valence fluctuation diverges 
$\chi_v=\infty$. The enegy gap opens for $\varepsilon_f>\varepsilon_f^{QCP}$. 
For $U_{fd}<U_{fd}^{QCP}$, the valence crossover occurs. 
At $U_{fd}=3.40$, $\bar{n}_f$ decreases continuously as $\varepsilon_f$ increases, as shown in Fig.~\ref{fig:nf_gap_Dlt}(c). 
At $\varepsilon_f=-28.80\equiv\varepsilon_f^{vc}$ which is defined as the valence-crossover point, $\chi_v$ has a peak indicating the enhanced valence fluctuation. The energy gap opens for $\varepsilon_f>-27.75$. 

\begin{figure}[t]
\includegraphics[width=8cm]{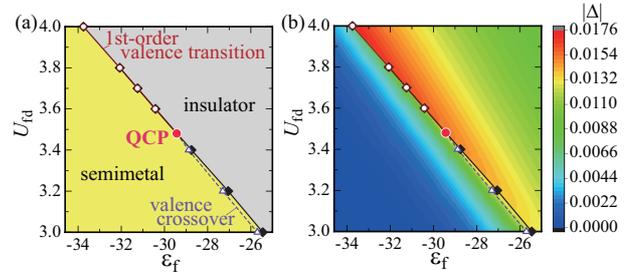}%
\caption{(color online) 
(a) Ground-state phase diagram in the plane of $f$-level and $4f$-$5d$ Coulomb repulsion. 
The FOVT line (open circle) terminates at the QCP (filled circle) from which the valence-crossover line extends (triangle with the dashed line). 
The MIT line (filled diamond) separates the semimetal phase and insulator phase. 
(b) Contour plot of the 
{EC}
parameter $|\Delta|$ at $T=0$. 
}
\label{fig:PD}
\end{figure}

We now discuss the effect of the exciton condensation. 
The magnitude of the 
{EC}
parameter $|\Delta|$ is plotted in Fig.~\ref{fig:nf_gap_Dlt}(a) for $U_{fd}=3.60$. 
As $\varepsilon_f$ increases from deep $\varepsilon_f$, $|\Delta|$ increases since the hybridization $|\bar{\Delta}(\xi)|$ with $\bar{\Delta}(\xi)$ being 
$\bar{\Delta}(\xi)\equiv\frac{1}{N}\sum_{\bm k}e^{i{\bm k}\cdot{\bm r}_{\xi}}\langle d^{\dagger}_{{\bm k}
{xy}
\sigma}f_{{\bm k}
{\Gamma_8^{(1)}}
\eta}\rangle$ 
{with $(\sigma,\eta)$=$(\uparrow,-)$, $(\downarrow,+)$}
for the N.N. Sm sites and $\bar{b}=\sqrt{z}$ increase in Eq.~(\ref{eq:Dlt}) reflecting the change from the Kondo regime to the intermediate-valence regime. 
Here, $z$ is the renormalization factor. 
At $\varepsilon_f=\varepsilon^{v}_f$, $|\Delta|$ shows a sudden rise since both $|\bar{\Delta}(\xi)|$ and $\bar{b}$ jump into larger values reflecting the FOVT into the weakly-correlated state with wider band width. As further $\varepsilon_f$ increases from $\varepsilon^{v}_f$, $|\Delta|$ increases slightly and then decreases monotonically. 
This decrease is because $|\bar{\Delta}(\xi)|$ decreases as the renormalized $f$ level $-\tilde{\varepsilon}_f$ goes down away from the $d$ $(t_{2g})$ band in the electron picture [see Fig.~\ref{fig:Ek}(d)] while $\bar{b}$ increases in Eq.~(\ref{eq:Dlt}). 
For $U_{fd}<U^{QCP}_{fd}$, the jump in $|\Delta|$ vanishes to be continuous function of $\varepsilon_f$ [see Figs.~\ref{fig:nf_gap_Dlt}(b) and \ref{fig:nf_gap_Dlt}(c)].

From these results, 
the ground-state phase diagram is determined in Fig.~\ref{fig:PD}(a). 
The FOVT line terminates at the QCP (filled circle), from which the valence-crossover line (dashed line) extends. 
The MIT line coincides with the FOVT line for $U_{fd}>U_{fd}^{\rm QCP}$, while 
for $U_{fd}<U_{fd}^{QCP}$ the MIT line slightly deviates toward larger $\varepsilon_f$ side as $U_{fd}$ decreases. 
Notable is that 
the 
{EC}
parameter $|\Delta|$ develops along the FOVT line and the valence-crossover line and is enhanced in the insulator phase in Fig.~\ref{fig:PD}(b).

As pressure is applied to SmS, the $f$ level increases since negative N.N. S ions approach the $4f$ electron at Sm [see Fig.~\ref{fig:Ek}(b)]. 
Hence, in the hole picture, $\varepsilon_f$ decreases as pressure increases. 
Thus 
decreasing $\varepsilon_f$ from the insulator phase to the semimetal phase for $U_{fd}>U_{fd}^{QCP}$ in Fig.~\ref{fig:PD}(a) is considered to correspond to 
the pressure evolution from $b$-SmS to $g$-SmS. 

\begin{figure}[t]
\includegraphics[width=6.5cm]{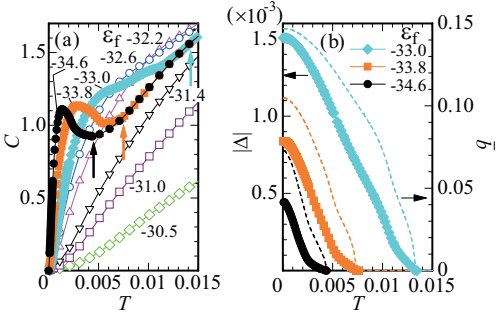}%
\caption{(color online) (a) Temperature dependence of the specific heat for each $\varepsilon_f$ at $U_{fd}=3.60$. 
Arrow indicates $T_c$ at which $\bar{b}$ starts to have a non-zero value. 
(b) Temperature dependence of $|\Delta|$ (left axis) and $\bar{b}$ (right axis) at $U_{fd}=3.60$. 
}
\label{fig:C_Dlt_y_T}
\end{figure}

To  
{get insight into}
anomalous behaviors in $g$-SmS under pressure, let us analyze the finite-temperature properties of the semimetal phase for $\varepsilon_f<\varepsilon_f^{v}$ at $U_{fd}=3.60$ [see Fig.~\ref{fig:PD}(a)] as a typical case. Figure~\ref{fig:C_Dlt_y_T}(a) shows the temperature dependence of the specific heat $C(T)$. 
Notable is that a hump structure appears, which becomes prominent as $\varepsilon_f$ decreases. 
{Interestingly, we find that the EC parameter $|\Delta(T)|$ shows a steep increase in the temperature range where $C(T)$ has a peak as shown in Fig.~\ref{fig:C_Dlt_y_T}(b). $|\Delta(T)|$ starts to have a non-zero value at $T_c$ where $\bar{b}$ starts to have a finite value [see Eq.~(\ref{eq:Dlt}) and vertical arrow in Fig.~\ref{fig:C_Dlt_y_T}(a) indicates $T_c$]. The transition of $\bar{b}(T)$ at $T=T_c$ is an artifact of the slave-boson mean-field treatment and in reality it should be a crossover to the heavy-fermion formation on cooling. The steep increase in $|\Delta(T)|$ appears in the low-temperature region for $T<T_c$ and hence this effect is considered to be intrinsic. This point is, however, to be examined by future analysis beyond the mean-field theory. The peak structure of $C(T)$ becomes more remarkable as $\varepsilon_f$ decreases, as seen for $\varepsilon_f=-33.8$ and $-34.6$ in Fig.~\ref{fig:C_Dlt_y_T}(a).}

\begin{figure}[t]
\includegraphics[width=6.5cm]{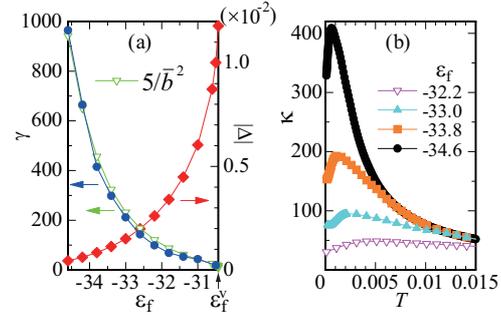}%
\caption{(color online) 
(a) The $\varepsilon_f$ dependence of $\gamma$ (circle, left axis), $5/\bar{b}$ (inverted triangle, left axis), and $|\Delta(T=0)|$ (diamond, right axis) at $U_{fd}=3.60$. 
(b) $\kappa$ vs $T$ at $U_{fd}=3.60$. 
}
\label{fig:comp_gamma}
\end{figure}

Figure~\ref{fig:comp_gamma}(a) shows the $\varepsilon_f$ dependence of the specific-heat coefficient $\gamma\equiv\lim_{T\to 0}C(T)/T$. 
As $\varepsilon_f$ decreases from $\varepsilon^{v}_f$, $\gamma$ increases with satisfying $\gamma\approx \bar{b}^{-2}=z^{-1}$. 
This indicates that the low-$T$ limit of the specific heat reflects the (semi)metallic ground state while the peak (hump) structure 
arises 
{at low temperatures.}
Namely, the ground state evolves into the Kondo regime as $\bar{n}_f$ approaches 1 while $|\Delta|$ decreases as $\varepsilon_f$ decreases. 

These features were observed in $g$-SmS under pressure~\cite{Matsubayashi}. 
In the high-pressure side of $g$-SmS, a Schottky peak appears in $C(T)$ ($P=10.1$, 15.9~kbar), whose temperature shifts to the lower $T$ side as pressure increases. 
Thus increase in $\gamma$ and decrease in the pseudo gap estimated from the Schottky specific heat in $g$-SmS under pressure~\cite{Matsubayashi} are explained by 
{our results}
in the semimetallic band, as shown in Fig.~\ref{fig:comp_gamma}(a). 
{Then,  these results support the pseudo gap~\cite{Matsubayashi,Kang2015} but no real gap in $g$-SmS. }

We find that the compressibility $\kappa\equiv\partial\bar{n}/\partial\mu$ also shows a peak at low temperatures as shown in Fig.~\ref{fig:comp_gamma}(b). 
Emergence of the peak structure in $\kappa$ implies that total charge fluctuations are enhanced. 
Similarly to the specific heat, the peak in $\kappa(T)$ is enhanced as $\varepsilon_f$ decreases, whose temperature shifts to the lower-$T$ side. 
This also explains the temperature dependence of the compressibility in $g$-SmS under pressure~\cite{Imura2009}. 

{


We have clarified the role of the $f$-level shift as a key parameter to understand the effect of applying pressure to SmS. It should be noted that applying pressure also enhances the $d$-$f$ hybridization. This effect brings the tendency to make the Sm valence $\nu=2+\bar{n}_f$ slightly smaller than the values obtained by the $f$-level shift only. Here we list $\nu$ at the FOVT for $\varepsilon_f\to\varepsilon_{f-}^{v} (\varepsilon_{f+}^{v})$ as 2.69 (2.44) at $U_{fd}=3.60$ and 2.89 (2.29) at $U_{fd}=4.00$. As $U_{fd}$ increases, the magnitude of the valence jump increases. 
Experimentally, in $b$-SmS, the Sm valence was observed as $\nu\sim 2$ with error bar about $\sim 0.2$ at the boundary of the 
{golden-black}
phase transition for $T=4.5$~K~\cite{Deen}. The gradual rise from $\nu\sim 2$ in $b$-SmS with increasing pressure was also detected by the M\"{o}ssbauer measurement~\cite{Coey1975}.  Recently, experimental technique for evaluating $U_{fd}$ directly in the rare-earth compounds has been developed in the X-ray spectroscopy measurement~\cite{Tonai}. Direct evaluation of the $U_{fd}$ value in SmS by this method is interesting future subject. 

{We have studied the isotropic condensation of the exciton as a first step analysis. It is interesting to examine whether the symmetry is not broken by exciton condensation. Detailed analysis of the crystal symmetry in $g$-SmS and also in $b$-SmS at low temperatures is interesting future subject. }

To summarize, we have clarified the mechanism of the interplay of the FOVT as well as MIT and exciton condensation in SmS.
Our study provides a unified view to resolve the long-standing issues in SmS. Our formulation can be generalized to the other Sm-based systems and hence this study opens a new stage for the quantum valence criticality, which offers the guide to explore the valence QCP in Sm-based compounds experimentally.

\begin{acknowledgment}
This work was supported by JSPS KAKENHI Grant Numbers JP18K03542 and JP19H00648.
\end{acknowledgment}


\end{document}